\documentclass[sigconf]{acmart}
\usepackage{xcolor}
\usepackage{multirow}
\usepackage{colortbl}
\usepackage{tabularx}
\newcolumntype{Y}{>{\centering\arraybackslash}X}
\definecolor{MistyRose}{rgb}{0.99, 0.91, 0.95}
\definecolor{myyellow}{rgb}{1,0.96,0.56}
\AtBeginDocument{%
  \providecommand\BibTeX{{%
    \normalfont B\kern-0.5em{\scshape i\kern-0.25em b}\kern-0.8em\TeX}}}

\setcopyright{none}
\renewcommand\footnotetextcopyrightpermission[1]{}
\def\ouralg{\textsc{POSO}}

\begin{document}


\title{\ouralg: Personalized Cold Start Modules for Large-scale Recommender Systems}


\author{Shangfeng Dai*, Haobin Lin*, Zhichen Zhao*, Jianying Lin, Honghuan Wu, Zhe Wang, Sen Yang, Ji Liu}
\email{{daishangfeng, linhaobin, zhaozhichen, linjianying, wuhonghuan, wangzhe, senyang, jiliu}@kuaishou.com}
\affiliation{%
  \institution{Kuaishou Technology}
  \city{Beijing}
  \country{China}
}

\newcommand\blfootnote[1]{%
  \begingroup
  \renewcommand\thefootnote{}\footnote{#1}%
  \addtocounter{footnote}{-1}%
  \endgroup
}


\begin{abstract}
  Recommendation for new users, also called user cold start, has been a well-recognized challenge for online recommender systems.
  Most existing methods view the crux as the lack of initial data.
  However, in this paper, we argue that there are neglected problems: 1) New users' behaviour follows different distributions from regular users.
  2) Although personalized features are used, heavily imbalanced samples prevent the model from balancing new/regular user distributions, as if the personalized features are overwhelmed.
  We name this problem as the ``submergence" of personalization. 
  To tackle this problem, we propose a novel method: \textbf{P}ersonalized C\textbf{O}ld \textbf{S}tart M\textbf{O}dules (POSO). From a model architecture perspective, POSO introduces user-group-specialized sub-modules to reinforce the personalization of existing modules.
   Then, it fuses their outputs by personalized gates, resulting in more comprehensive representations.
  In this way,
  POSO equalizes imbalanced features by assigning individual combinations of modules.
  POSO can be flexibly integrated into many existing modules such as Multi-layer Perceptron, Multi-head Attention and Multi-gated Mixture of Experts, and effectively improves their performance with negligible computational overheads.
  The proposed POSO shows remarkable advantage in large-scale industrial  recommendation scenario.
  It  has been deployed on Kwai and improves new user Watch Time by a large margin (+7.75\%). Moreover, POSO can be further generalized to regular users, inactive users and returning users (+2\%-3\% on Watch Time), as well as item cold start (+3.8\% on Watch Time).
  Its effectiveness has also been verified on public dataset (MovieLens 20M).
  We believe the proposed POSO can be well generalized to other scenarios.
\end{abstract}





\maketitle
\section{Introduction}

\blfootnote{*equal contribution}

\begin{figure}[h]
  \centering
  \includegraphics[width=0.8\linewidth]{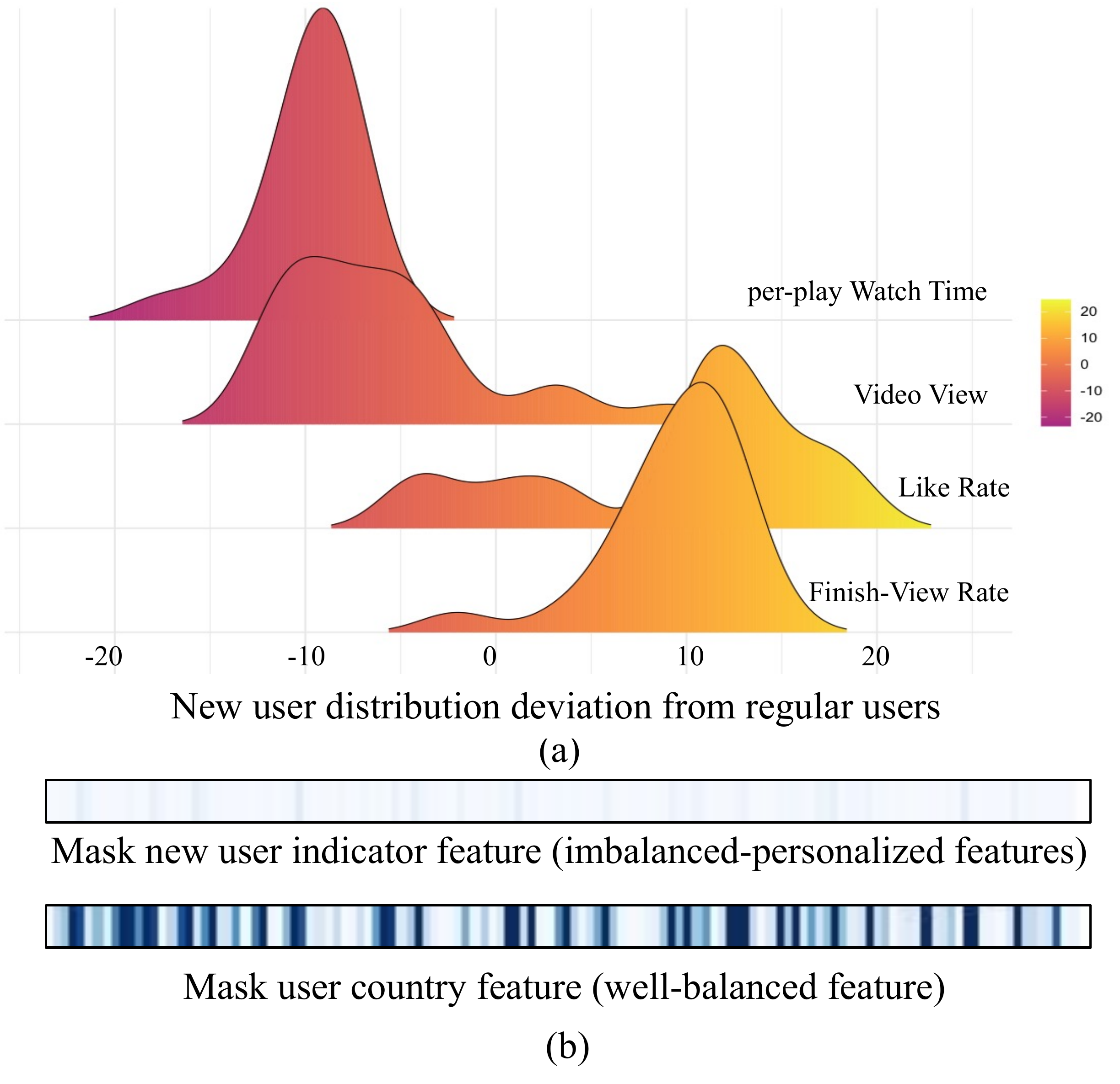}
  \caption{(a) Visualization of new user posterior behavior (based on the relative difference of action count/rate from regular users).
  It shows that new users follow very different distributions from regular users. (b) Sensitivity of imbalanced and well-balanced features,  visualized by two vectors of size 128. Bins in each vector present activation difference when masking imbalanced/balanced features. Deeper color visualizes more significant difference.}
  \label{fig:statistic}
\end{figure}
\begin{figure}[h]
  \centering
  \includegraphics[width=0.8\linewidth]{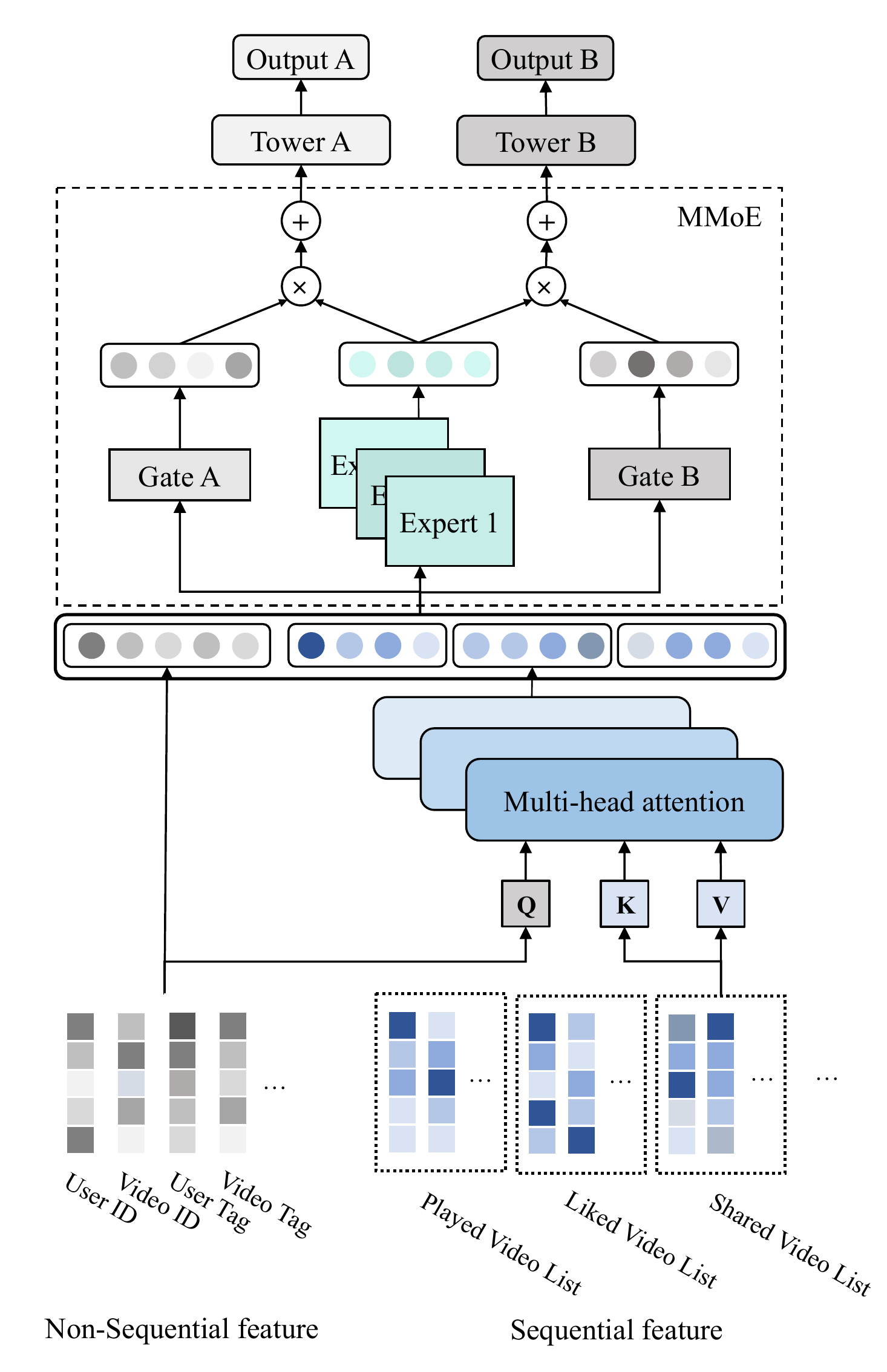}
  \caption{Existing model on the large-scale recommender system of Kwai. There are 3 stages: embedding generation, sequential feature modeling (by MHA) and multi-task optimizing (by MMoE).}
  \label{basemodel}
\end{figure}

Large-scale Recommender Systems (RS) confronts numerous new visitors everyday.
One challenging but important problem is how to make accurate recommendation for these unseen users. On one hand, these users hardly have historical description or initial data. On the other hand, they are more sensitive and impatient than regular users. Inaccurate recommendation fails to draw their attention, prompting them not to return the platform. Then we probably lose the potential value of new users.

The problem is known as ``user cold start problem"~\cite{usercoldstart1,usercoldstart2}. Unlike item cold start problem~\cite{itemcoldstart} where we can exploit
content features~\cite{imagematters,AAAI1816066,whatyoulookmatters}, user cold start hardly provides alternative description and requires the system to swiftly capture user interests.
Meta-learning based methods~\cite{melu,metae} mitigate the problem by producing well generalized initialization. Besides, other work~\cite{learning2hash,mwuf} tries to generate ID embedding by the rest features, thus supplying missing cues.

However, we argue that there exists another neglected problem: the submergence of personalization. The problem describes a phenomenon that even though personalized features are used to balance various user groups (whose distributions are very different), 
these features are overwhelmed because of heavily imbalanced samples.

As shown in Fig.\ref{fig:statistic} (a), we average regular users' posterior behaviour (Watch Time/Video View Count/Like Rate/Finish-View Rate) as original points, and show distributions of new users. It is shown that new users follow very different distributions. 
Theoretically, we expect personalized features to distinguish user groups. Do such features help the model balance various distributions in practice? The answer is \textbf{NO}. We find that the personalized inputs are overwhelmed, as shown in Fig.\ref{fig:statistic} (b).
In both cases we use the same well-trained model, and visualize the activation difference (near the end of the network, averaged across multiple batches) when some features are masked by $0$. In the former one we mask new user indicator (0 for regular users and 1 for new users). Surprisingly, the activations almost keep unchanged. The reason is that such features are heavily imbalanced: new users' samples only occupy less than 5\% over all samples. In the training procedure the indicator barely changes most of the time, so this feature becomes dispensable.
In contrast, in the latter one a well-balanced feature (user country) is masked. 
Unlike the former case, activations significantly change. 
The above observation suggests that a plain model architecture is insufficient to maintain personalization.

In this paper, 
we propose an effective module to solve the above problems: \textbf{P}ersonalized C\textbf{O}ld \textbf{S}tart M\textbf{O}dules (POSO). First, POSO equalizes imbalanced samples by assigning individual combinations of modules, each of which only focuses on its assigned user groups.
Then, POSO generates personalized gates varying from raw personalized features. At last gate and module outputs are combined to form comprehensive representations.
Its effectiveness is two-fold: 1) Samples are evenly assigned to specialized sub-modules, regardless of majority or minority. 2) Gating network is fully determined by the chosen personalized features (called ``Personalization Code"), which avoids their ``submergence".
POSO reinforces personalization, balancing various distributions and mitigating the cold start problem.
POSO does not serve as a standalone method. It can be integrated into many existing modules, such as Multi-layer Perceptron (MLP), Multi-head Attention (MHA) and Multi-gated Mixture of Experts (MMoE). By proper approximation and detailed analysis, we derive their personalized versions, which brings compelling gains consistently but with negligible computational overheads. 

One of the advantages of POSO is that it excellently benefits large-scale systems: 1) It follows the standard training procedure, unlike meta-learning based methods which manually split training data into support/query set and probably slow down training speed. 2) The computational overheads are negligible. 3) It can be adopted to other data imbalance problems, which widely exists in users/items/countries/regions.

We conduct extensive experiments on large-scale recommender system of Kwai as well as public dataset. In real-world scenarios, 
POSO (MLP)/POSO (MHA)/POSO (MMoE) consistently improve the performance, and outperform existing methods.
When deployed to our online system, it brings +7.75\% Watch Time and +1.52\% Retention Rate for new users. Meanwhile, it benefits regular/inactive/returning users (+2-3\% Watch Time) as well.
Besides user cold start scenario, the proposed architecture improves item cold start (+3.8\% Watch Time for new videos) and outperforms existing methods on MovieLens 20M dataset~\cite{movielens}.

The contributions of this paper are summarized as follows: 
\begin{enumerate}
\item We reveal the submergence of personalization problem. Without customized architectures, personalized features can be overwhelmed. That eventually hurts the performance.
\item We propose a novel method named POSO, which reinforces personalization under imbalance data, and significantly mitigates the cold start problem.
\item We present detailed derivation and show that POSO can be integrated into many existing modules with negligible computational overheads. The personalized modules advance the large-scale recommender system by a large margin.
\end{enumerate}

\section{Related Work}
Related research for user cold start problem can be summarized into two genres: meta-learning and embedding generation.
Meta-learning refers to a series of methods aiming at training generalized networks to produce well predictions for brand-new tasks~\cite{metalearning,metalearningsurvey}.
MAML~\cite{maml} shows promising results on few-shot learning, but mainly focuses on classification tasks. Following its idea, meta-learning based methods are introduced to the recommender system: 
MeLU~\cite{melu} treats recommendation for each user as an individual task. In local update steps, embedding receives no gradients to ensure the stability of the network. Similarly, 
Du and Wang et al.~\cite{scenariometa} use meta-learning to transfer knowledge between scenarios, e.g. from traveling
task to babysitting task.
The work of ~\cite{ametalearning} successfully implements meta-learning strategy on production data. It has two architectures to adjust weights in Matrix Factorization methods. DropoutNet~\cite{dropoutnet} can be viewed as a similar try on improving generation. It randomly masks user inputs to imitate new users.


Another genre tries to generate meaningful IDs embedding by other features. 
Meta-E~\cite{metae} learns to generate user IDs embedding from other embedding.
The learning procedure is supervised by cold-start phase and warm-up phase respectively. 
MAMO~\cite{mamo} proposes multiple memory: profile memory, user memory and task-specific memory. Such memory can be viewed as user feature bases, which are used to decompose cold users into warm features. 
MWUF~\cite{mwuf} believes that there exists scaling difference between regular/new items for item embedding and shifting difference for users. The final embedding is formed by scaling and shifting networks.

\section{Existing Production Model}\label{sec:basemodel}

In this section, we briefly describe the structure of the existing production model on the large-scale recommender system of Kwai.
As illustrated in Fig.\ref{basemodel}, the model follows the classic Embedding\&MLP paradigm~\cite{resembedding}.
In addition, some advanced modules (e.g. MHA, MMoE) are introduced to achieve better performance. 

The inputs are composed of non-sequential features (e.g. user ID) and sequential features (e.g. user's past watched videos). In the embedding generation stage, all features are firstly mapped into low dimensional vectors, by an embedding look-up table. For each sequential feature, a Multi-Head Attention (MHA) module~\cite{transformer} is applied to fuse the sequence of embedding into a single one, which has been introduced to recommender system by ~\cite{bst}. 
In existing implementation, keys ($K$) and values ($V$) are produced by linear projection of sequential embedding $X^{\text{seq}}$. Namely, $K = W^K X^{\text{seq}}$, and $V = W^V X^{\text{seq}}$ ($W^Q, W^K$ and $W^V$ are trainable matrices)\footnote{In this paper, superscript names a specific modules. Signs like $(i)$ indexes modules.
While $[\cdot]_i$ index the i-th element of a vector}.
Differently, query $Q$ receives concatenated non-sequential embedding $\mathbf{x}^{\text{non}}$ as inputs: $Q = W^Q$ $\mathbf{x}^{\text{non}}$. For a single head, 
$\text{head}(Q, K, V) = \text{softmax}\left(\frac{QK^\top}{\sqrt{d^h}}\right)V$, 
 where $d^h$ denotes the dimension of projected features. 
 The result of MHA is simply the concatenation of each head output.

In the following stage, all non-sequential embedding and transformed sequential embedding is concatenated as intermediate activation $\mathbf{x}$.
The production model needs to predict $T$ targets simultaneously, such as Long-View Rate and Like Rate (see the definition in Sec.\ref{sec:offline}). In order to model the task relationships, the Multi-gate Mixture of Experts (MMoE) module~\cite{mmoe} is introduced, with $N^e$ MLPs $\{e^{(i)}\}_{i=1}^{N^e}$ as experts. A gating network $g^t$ is trained for task $t$, which ensembles the expert outputs into $\mathbf{\hat{x}}^t$. Finally, a task-specific MLP $h^t$ takes $\mathbf{\hat{x}}^t$ and gives the prediction $y^t$. The formulation of the MMoE module is given by:
\begin{equation}
\begin{aligned}
g^t(\mathbf{x}) = \text{softmax}\left( W^{t}\mathbf{x} \right),  
\label{eq:mmoe1}
\end{aligned}
\end{equation}
\begin{equation}
\begin{aligned}
\mathbf{\hat{x}}^t = \sum_{i=1}^{N^e}\left[g^t\left(\mathbf{x}\right)\right]_ie^{(i)}\left(\mathbf{x}\right), 
\label{eq:mmoe2}
\end{aligned}
\end{equation}
\begin{equation}
\begin{aligned}
y^t = h^t\left(\mathbf{\hat{x}}^t\right), 
\label{eq:mmoe3}
\end{aligned}
\end{equation}
for $t=1, 2, ..., T$, where $W^{t}$ is the trainable matrix for gating network.

\begin{figure*}[h]
  \centering
  \includegraphics[width=0.9\linewidth]{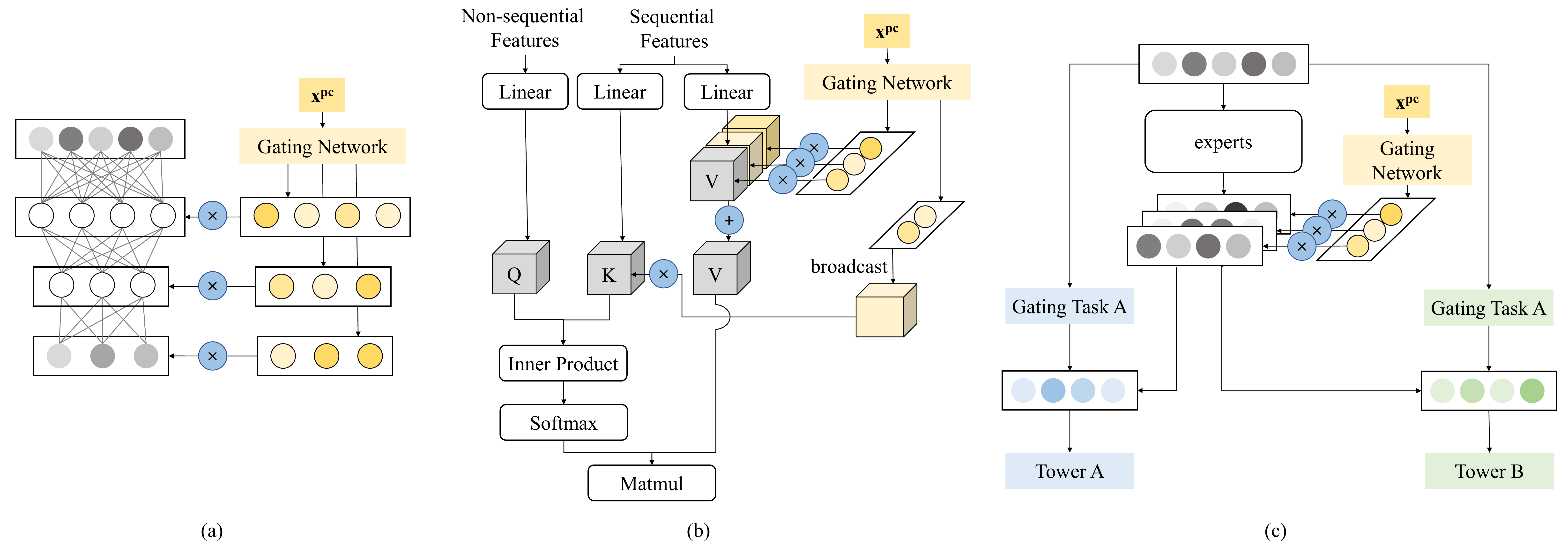}
  \caption{Personalized modules by POSO: (a) POSO (MLP) masks each activation in each layer respectively. (b) In POSO (MHA), $Q$ is not personalized, $K$ is lightly personalized and $V$ is totally personalized. (c) In POSO (MMoE), personalization is firstly adopted, then the output fed the specific tasks. All modules of POSO are colored by yellow.}
  \label{fig:pmoms}
\end{figure*}

\begin{table}
\centering
\begin{tabular}{ccc}
\toprule

Method & \#Params & \#FLOPs \\
\midrule
MLP & 2.48M & 1.50M \\
POSO (MLP) & 2.59M & 1.85M \\
\hline
MHA & 2.91M & 7.82M \\
POSO (MHA) & 3.15M & 8.55M \\
\hline 
MMoE & 13.20M & 40.10M \\
POSO (MMoE) & 13.20M & 40.10M \\
\hline 
Baseline & 23.15M & 69.46M \\
Overall (All Combined) & 23.51M & 70.54M \\
\bottomrule
\end{tabular}
\caption{The comparison on params and FLOPs of the production model.}
\label{tab:complexity}
\end{table} 

\section{Personalized Cold Start Modules}\label{sec:pmom}
It is well-known that the system suffers from lack of initial data for new users. However, we argue that one problem has been neglected: the ``submergence" of personalization, which means the system fails to balance various distributions because of data imbalance, even though personalized features are provided.

First, we show that new users' behaviour distributes differently with regular users'. In Fig.\ref{fig:statistic} (a) we visualize posterior behaviour of new/regular users. Metrics of regular users are averaged as original points. We show the relative difference of new user metrics. We observe that 1) New users produce lower Video View (VV), indicating that
the system hardly captures their interests. 2) New users have higher Finish-View Rate but lower per-play Watch Time. They
may enjoy short videos, but have little patience on long videos. 3) New users tend to ``like" more frequently, seeming to feel fresh for a wide range of videos.
All the observations imply that new users' behaviour follows very different distributions from regular users.

One may believe that existing model implicitly balances various distributions by utilizing personalized features for granted, such as an indicator that distinguishes new/regular users.
However, because of data imbalance, such features are overwhelmed. In Fig.\ref{fig:statistic} (b) we utilize a well-trained model, mask personalized features and visualize the activation difference. Surprisingly, masking heavily imbalanced new user indicator almost has no impact on activation. On the contrary, when masking well-balanced user country feature, the activation significantly changes. 
Since new users only occupy 5\% samples. Most of the time, the indicator keeps unchanged. The model easily concentrates on other features to search for solutions, and ``forget" the new user indicator, which is critical for cold start problem.
We call such problem as the ``submergence" of personalization.

In this paper, we reinforce personalized features from a model architecture perspective. 
We equalize imbalanced-personalized features by assigning individual combinations of models to solve the submergence problem.
Ideally, one can construct an exclusive model for a specific user:
\begin{equation}
\begin{aligned}
y^{u}=f^{u}\left(\mathbf{x}^u\right),
\label{eq:origin}
\end{aligned}
\end{equation}
where $\mathbf{x}$, $y$, $f$ denote inputs, outputs and the model respectively. Subscript ``$u$" refers to a specific user. In such scheme personalization is fully preserved in the corresponding models.
Unfortunately, due to the large amount of users, the proposal above is infeasible. One possible solution is establishing several individual models for each kind of user groups (such as new users, returning users and so on). One specific user can be viewed as combination of various user groups (e.g. one can be half inactive user and half regular user).
Subsequently, we can decompose the prediction for the specific user as combination of predictions for user groups:
\begin{equation}
\begin{aligned}
y^u=\sum_{i=1}^N w_i f^{\left(i\right)}\left(\mathbf{x}\right),
\end{aligned}
\end{equation}
where $i$ denotes model index and we have $N$ models.
In practice, it is hard to generate $w_i$. Instead, we use gating networks to produce $w_i$ from personalized features: $w_i=\left[g(\mathbf{x}^{\text{pc}})\right]_i$ where pc refers to Personalization Code, i.e. critical features that identify user group. 
So far, we still have to prepare $N$ individual models for capturing user group interests, which is computationally intensive. 
One key point of our method is that we apply decomposition on intermediate modules, and maintain the rest modules unchanged:
\begin{equation}
\begin{aligned}
\mathbf{\hat{x}}=C\sum_i^{N}\left[g\left(\mathbf{x}^{\text{pc}}\right)\right]_i f^{\left(i\right)}\left(\mathbf{x}\right),
\label{eq:pmom-final}
\end{aligned}
\end{equation}
where $f$ denotes modules from now, $\mathbf{\hat{x}}$ and $\mathbf{x}$ are activations of two adjacent layers.
Note there is no constraints on the sum of $g(\mathbf{x})$, to avoid overall scale drifting, a rectified factor $C$ is applied.

Eq.\ref{eq:pmom-final} shows the prototype of the proposed method. As it introduces personalization into intermediate modules, we name it ``\textbf{P}ersonalized C\textbf{O}ld \textbf{S}tart M\textbf{O}dules (POSO)".

The design of POSO is marked by the following principles:

\noindent \textbf{Personalization}. 
The POSO solves the submergence in two aspects: 1) Features are equalized by assigning multiple modules and gates. 
Even though regular user data dominates, new user samples wouldn't be neglected because POSO exploits another set of modules and gates to make predictions.
2) Whichever layer to be applied, POSO highlights personalization through raw features instead of second-hand activations, which is hardly achieved by self-learning techniques (such as MoE, see Sec.\ref{sec:gates}).

\noindent \textbf{Flexibility}. 
Note that POSO is not a standalone module, but a general approach to personalize existing modules. POSO can be integrated into many existing methods, and equips them with personalization. In the following, we derive personalized versions of MLP, MHA and MMoE.
We also believe it has favorable prospect when applied on other unexplored modules. 

\noindent \textbf{Non-aftereffect}. 
Sub-modules of POSO share the same input and their outputs are finally fused into single comprehensive result. This ensures the structural independency. No dependency is introduced between upstream and downstream modules.


\subsection{POSO of Linear Transformation}\label{sec:pmom-linear}
We begin with the most basic module: linear transformation, which is formulated as $f(\mathbf{x}) = W\mathbf{x}$,
where $\mathbf{x}\in R^{d^{\text{in}}}$ and $\mathbf{\hat{x}}\in R^{d^{\text{out}}}$. Substituting its formulation into Eq.\ref{eq:pmom-final} gives
\begin{equation}
\begin{aligned}
\mathbf{\hat{x}} &= C\sum_{i=1}^N \left[g\left(\mathbf{x}^{\text{pc}}\right)\right]_i W^{\left(i\right)}\mathbf{x}.
\label{eq:pmom-li-mid}
\end{aligned}
\end{equation}
Specifically, the $p$-th entry of $\mathbf{\hat{x}}$ is given by
\begin{equation}
\begin{aligned}
\mathbf{\hat{x}}_p = C\sum_{i=1}^N\sum_{q=1}^{d^{\text{in}}} \left[g\left(\mathbf{x}^{\text{pc}}\right)\right]_iW^{\left(i\right)}_{p,q} \mathbf{x}_q,
\label{eq:pmom-li}
\end{aligned}
\end{equation}
where $W^{(i)}_{p.q}$ refers to the element of $W^{(i)}$ at location $(p,q)$.
Though Eq.\ref{eq:pmom-li} introduces $N$ times of complexity, sufficient free parameters allow us to apply simplification in flexible ways. Here we present a simple but effective case.
Let $N=d^{\text{out}}$, $W^{(i)}_{p, q}=W_{p, q} $ $\forall p, q$ when $i=p$, and $W^{(i)}_{p, q}\equiv 0$ for any $i \ne p$. We have:
\begin{equation}
\begin{aligned}
\mathbf{\hat{x}}_p = C \cdot \left[g\left(\mathbf{x}^{\text{pc}}\right)\right]_p \sum_{q=1}^{d^{\text{in}}} W_{p,q}\mathbf{x}_q,
\label{eq:pmom-li-entry}
\end{aligned}
\end{equation}
or equivalently,
\begin{equation}
\begin{aligned}
\mathbf{\hat{x}} = C \cdot g\left(\mathbf{x}^{\text{pc}}\right) \odot W \mathbf{x},
\label{eq:pmom-li-final}
\end{aligned}
\end{equation}
where $\odot$ denotes element-wise multiplication.
This simplification leads to a computationally efficient operation: just applying element-wise multiplication on the original output by the personalized gates. 

\subsection{POSO of Multi-Layer Perceptron}
Following the similar derivation in Sec.\ref{sec:pmom-linear}, the personalized version of Fully-Connected layer (FC) with activation function is designed as:
\begin{equation}
\begin{aligned}
\mathbf{\hat{x}} = C \cdot g\left(\mathbf{x}^{\text{pc}}\right)\odot \sigma \left(W\mathbf{x}\right),
\label{eq:pmom-fc-final}
\end{aligned}
\end{equation}
where $\sigma$ denotes activation function.
It presents a similar form with LHUC~\cite{lhuc}, whose hidden unit contributions are here replaced by personalized gates.

Naturally, the personalized version of MLPs, called POSO (MLP), is formed by stacking the personalized FCs. Its framework is shown in Fig.\ref{fig:pmoms} (a). In Table.\ref{tab:complexity} we detail params and FLOPs of each module, and figure that the proposed modules are computationally efficient.

\subsection{POSO of Multi-Head Attention}
In this part, we derive POSO version of the Multi-Head Attention (MHA) module.
For clarity, let's first consider the formulation of a single head:
\begin{equation}
\begin{aligned}
\mathbf{\hat{x}} = \text{softmax}\left(\frac{QK^\top}{\sqrt{d^h}}\right)V.
\label{eq:mha}
\end{aligned}
\end{equation}
By substituting Eq.\ref{eq:mha} into Eq.\ref{eq:pmom-final} as $f^{(i)}$ we have:
\begin{equation}
\begin{aligned}
\mathbf{\hat{x}}=C\sum_{i=1}^{N}\left[g\left(\mathbf{x}^{\text{pc}}\right)\right]_i\left(\text{softmax}\left(\frac{Q^{\left(i\right)}\left(K^{\left(i\right)}\right)^\top}{\sqrt{d^h}}\right)V^{\left(i\right)}\right).
\label{eq:pmom-mha}
\end{aligned}
\end{equation}
This naive implementation introduces multi-fold $Q$, $K$ and $V$. Despite improving the performance (see Sec.\ref{sec:degree}), it is computationally expensive.
To reduce overheads, we rethink the role of $Q$, $K$ and $V$. 

Firstly, $Q$ contains all user features except historical behaviour, so it's already highly personalized. Therefore, we just set $Q^{(i)}=Q, \forall i$.
On the contrary, $V^{(i)}$ involves little user information. Considering that $V$ directly determines output, we take no simplification on multi-fold $V^{(i)}$.
We notice that using multi-fold $K$ introduces redundant free parameters, since the attention weight produced by $K$ and $Q$ has much lower dimension than $K$ itself. Alternatively, a personalized gate $G^k$ for element-wise multiplication as introduced in Sec.\ref{sec:pmom-linear} is sufficient to adjust attention weight, i.e. $K^{(i)}=G^k(\mathbf{x}^{\text{pc}})\odot K$.
\footnote{In fact $G^k(\mathbf{x}^{\text{pc}})$ is a 2-dimensional tensor while $K$ is a 3-dimensional (including the batch dimension) tensor, so the element-wise operation broadcasts on the last dimension of $K$.}.

By now, both $Q$ and $K$ become irrelevant to $i$ and thus can be moved out from the summation.
Eq.\ref{eq:pmom-mha} is then simplified as:
\begin{equation}
\begin{aligned}
\mathbf{\hat{x}} =C \cdot \text{softmax}\left(\frac{Q \cdot \left(G^k\left(\mathbf{x}^{\text{pc}}\right)\odot K\right)^\top}{\sqrt{d^h}}\right)\sum_{i=1}^{N}\left[g(\mathbf{x}^{\text{pc}})\right]_iV^{\left(i\right)}.
\label{eq:pmom-mha-mid}
\end{aligned}
\end{equation}

In summary, we respectively personalize the components at 3 levels: no personalization for $Q$, light-weight personalization for $K$ and full personalization for $V$. The extent of personalization on these three tensors also agrees with their roles in MHA, as mentioned above. 
Finally, for multi-head cases, outputs of each head are concatenated to form the representations.

The proposed module is named as ``POSO (MHA)", whose framework is shown in Fig.\ref{fig:pmoms} (b). In our scenario, compared with the original version of MHA, POSO (MHA) has comparable complexity (see Table.\ref{tab:complexity}) but significantly better performance (see Sec.\ref{sec:degree}).

\begin{table*}
\centering
\begin{tabular}{c|cccc|cccc}
\toprule
 &
  \multicolumn{4}{c|}{New Users} &
  \multicolumn{4}{c}{Regular Users} \\ \cline{2-9} 
\multirow{-2}{*}{Methods} &
  \multicolumn{1}{c}{Long-View} &
  \multicolumn{1}{c}{Finish-View} &
  \multicolumn{1}{c}{Like} &
  \multicolumn{1}{c|}{Follow} &
  \multicolumn{1}{c}{Long-View} &
  \multicolumn{1}{c}{Finish-View} &
  \multicolumn{1}{c}{Like} &
  \multicolumn{1}{c}{Follow} \\ \hline
MeLU~\cite{melu} &
  \cellcolor[HTML]{FFF0F5}-0.135 &
  \cellcolor[HTML]{FFF0F5}-0.099 &
  +0.124 &
  +0.160 &
  \cellcolor[HTML]{FFF0F5}-0.068 &
  \cellcolor[HTML]{FFF0F5}-0.038 &
  +0.028 &
  -0.093 \\
Meta-E~\cite{metae} &
  \cellcolor[HTML]{FFF0F5}-0.024 &
  \cellcolor[HTML]{FFF0F5}+0.024 &
  -0.260 &
  -0.338 &
  \cellcolor[HTML]{FFF0F5}-0.038 &
  \cellcolor[HTML]{FFF0F5}+0.005 &
  -0.283 &
  -0.345 \\
MWUF~\cite{mwuf} &
  \cellcolor[HTML]{FFF0F5}+0.023 &
  \cellcolor[HTML]{FFF0F5}-0.002 &
  +0.392 &
  +0.123 &
  \cellcolor[HTML]{FFF0F5}+0.072 &
  \cellcolor[HTML]{FFF0F5}+0.032 &
  +0.212 &
  +0.136 \\ \hline
POSO (MLP) &
  \cellcolor[HTML]{FFF0F5}+0.188 &
  \cellcolor[HTML]{FFF0F5}+0.121 &
  +0.314 &
  \textbf{0.761} &
  \cellcolor[HTML]{FFF0F5}+0.277 &
  \cellcolor[HTML]{FFF0F5}\textbf{+0.173} &
  +0.095 &
  +0.775 \\
POSO (MHA) &
  \cellcolor[HTML]{FFF0F5}+0.279 &
  \cellcolor[HTML]{FFF0F5}+0.190 &
  +0.089 &
  +0.529 &
  \cellcolor[HTML]{FFF0F5}+0.251 &
  \cellcolor[HTML]{FFF0F5}+0.150 &
  -0.008 &
  \textbf{+0.892} \\
POSO (MMoE) &
  \cellcolor[HTML]{FFF0F5}+0.223 &
  \cellcolor[HTML]{FFF0F5}+0.132 &
  \textbf{+0.428} &
  +0.388 &
  \cellcolor[HTML]{FFF0F5}+0.295 &
  \cellcolor[HTML]{FFF0F5}+0.162 &
  +0.184 &
  +0.726 \\
POSO (All Combined) &
  \cellcolor[HTML]{FFF0F5}\textbf{+0.442} &
  \cellcolor[HTML]{FFF0F5}\textbf{+0.248} &
  +0.344 &
  +0.211 &
  \cellcolor[HTML]{FFF0F5}\textbf{+0.339} &
  \cellcolor[HTML]{FFF0F5}+0.171 &
  \textbf{+0.329} &
  +0.492 \\ \bottomrule
\end{tabular}
\caption{Results (percent point) of offline experiments, compared with baseline. Colored tasks are more important.}
\label{tab:offline_newuser}
\end{table*}

\subsection{POSO of Multi-gated Mixture of Experts}

In this part, we present the POSO version of MMoE. 
Substituting Eq.\ref{eq:mmoe2} into Eq.\ref{eq:pmom-final} as $f^{(i)}$ gives:
\begin{equation}
\begin{aligned}
\mathbf{\hat{x}}^t = C\sum_{i=1}^{N}\left[g\left(\mathbf{x}^{\text{pc}}\right)\right]_i\left(\sum_{j}^{N^e}\left[g^t\left(\mathbf{x}\right)\right]_je^{\left(j\right)}\left(\mathbf{x}\right)\right), \\
\end{aligned}
\label{eq:pmom-mmoe}
\end{equation}
where $i, j, t$ index personalized gates, experts and tasks.
In Eq.\ref{eq:pmom-mmoe} there are two implicit constraints: each group of experts shares the same personalized gate $g^{(i)}$, each group of $g^t$ is normalized by Softmax. We relax the constraints to simplify the implementation. First, we allow each expert to have its own personalized gate. Then we implement normalization over all task gates. Thereby we have:
\begin{equation}
\begin{aligned}
\mathbf{\hat{x}}^t = C\sum_{i=1}^N\sum_{j=1}^{N^e}\left[g\left(\mathbf{x}^{\text{pc}}\right)\right]_{ij}\left[g^t\left(\mathbf{x}\right)\right]_{ij}e^{\left(ij\right)}\left(\mathbf{x}\right),
\end{aligned}
\label{eq:pmom-mmoe-mid}
\end{equation}
where $g^t$ is normalized over all pairs of $(i, j)$.
Note that in Eq.\ref{eq:pmom-mmoe-mid} the indices $i$ and $j$ jointly index experts. Let $\hat{N}=NN^e$, we can re-index the modules and rewrite the above equation as:
\begin{equation}
\begin{aligned}
\mathbf{\hat{x}}^t = C\sum_{i=1}^{\hat{N}}\left[g\left(\mathbf{x}^{\text{pc}}\right)\right]_i\left[g^t\left(\mathbf{x}\right)\right]_{i}e^{\left(i\right)}(\mathbf{x}),
\end{aligned}
\label{eq:pmom-mmoe-final}
\end{equation}
\begin{equation}
\begin{aligned}
g^t(\mathbf{x}) &= \text{softmax}\left(W^{t}\mathbf{x}\right).
\end{aligned}
\end{equation}

The overall unit count $\hat{N}$ is actually a hyper-parameter that can be manually adjusted. In our implementation we just set $\hat{N}=N$ to save computation complexity.

In Eq.\ref{eq:pmom-mmoe-final} we obtain the finalized version of personalized MMoE, namely, POSO (MMoE). 
The implementation is extremely light-weighted (also see Table.\ref{tab:complexity}): one can keep all the structure of MMoE, and just mask each expert by its personalized gate, as shown in Fig.\ref{fig:pmoms} (c).

How the POSO (MMoE) improves experts performance?
In MMoE, experts are only task-aware, but have ambiguous knowledge on samples. In POSO (MMoE), experts are personalized activated:
if samples belonging to new users produce higher weight in $g\left[\cdot\right]_i$, the corresponding i-th expert obtains higher learning weight, and becomes more sensitive for new users, vice versa. In such way 
experts become specialized. 
We can say the experts are not only task-aware, but also field-aware on user groups. 
In sec.\ref{sec:vis} we visualize gating network outputs of value matrices in MHA. They are similarly specialized.

\subsection{POSO for Cold Start}\label{sec:coldstart}
Now, we demonstrate how to mitigate the cold start problem, with the knowledge of POSO.

\noindent \textbf{User Cold Start}. New users are defined as users whose first launch happens within $T_{du}$ hours. 
For user cold start, we exploit a fine-grained feature to reveal how many items have been impressed for this user, i.e. bucketized Accumulated View Count (AVC). This feature is fed into the gating network $g$ as the PC. In each module, we keep the same input for gating network
and intensify personalization.

\noindent \textbf{Item Cold Start}. The definition of new item (video) is two-fold: 1) It is uploaded within $T_{dv}$ days and 2) Its overall impression count is less than $T_{s}$. Similarly, we exploit video age to distinguish regular/new video. It still produces personalization, but from the view of videos.

In this paper, the gating network is composed of a two-layer MLP, whose outputs are activated by Sigmoid functions.

\section{Experiments}
In this section, we present the performance of POSO on large-scale recommendation scenario of Kwai. We conduct both offline and online experiments. We also validate the generalization of POSO on public dataset. Besides, we demonstrate how to select PC and show visualization inspiration for personalized modules.

\subsection{Offline Experiments}\label{sec:offline}

\noindent \textbf{Dataset Setup}. For offline experiments, samples come from our large-scale recommender system. 
We build training dataset from 7-consecutive-day record and test dataset from the day next. 

\noindent \textbf{Tasks}. In video recommender system, users may have two kinds of feedbacks. Explicit feedbacks can be giving a like (in the following we use ``Like" to denote) and deciding to follow an author (denoted as ``Follow"). Implicit feedbacks mainly refer to whether the user watches the video long enough (Long-View) or completely finishes the video (Finish-View). We follow a multi-task framework to optimize Long-View/Finish-View/Like/Follow Rate simultaneously. Empirically,
Long-View and Finish-View are more authoritative metrics that determine online performance,
In this paper, a watching event is defined as ``Long-View" when its duration exceeds the $T_{lv}$ percent of the video length and thus it is modeled as a CTR-like task.

\noindent \textbf{Metrics}. 
In our experiments, we adopt GAUC~\cite{ocpc} to measure the performance of a model, where the AUC is first calculated within samples of each user, and averaged w.r.t sample count. 

To validate the effectiveness of the proposed method, we compare various POSO with the existing approaches that are also focusing on cold start problem.
MELU~\cite{melu} utilizes Meta Learning~\cite{maml} in recommender system, and formulates cold start problem as few-shot learning.
Meta-E~\cite{metae} and MWUF~\cite{mwuf} considers generate ID embedding to supply missing cues.
These methods cover various aspects from optimization to embedding initialization. 

The results are shown in Table.\ref{tab:offline_newuser}, where ``Rate" is omitted. Because of privacy policy, we only show the absolute difference between baseline and other methods, denoted as percent point (pp).
MeLU moderately improves Like and Follow Rate. However, it fails on watching tasks. It seems that MeLU performs well when the task has sparse positive samples. In Meta-E, the interaction tasks drop in turn.
MWUF provides improvement on regular users and interaction tasks for new users. 
Potentially its embedding can be supplementary for IDs embedding (such as user ID embedding, user Tag embedding). However, in industrial scenario, numerous features supply ID embedding, and the improvement is mostly covered.

The comparison of POSO (MLP), POSO (MHA) and POSO (MMoE) is interesting: All of them improve watching tasks. POSO (MHA) performs better on new users while POSO (MMoE) prefers regular users. Meanwhile, POSO (MLP) also focuses more on regular users.
It implies that primary heads of MHA redundantly focus on regular users while experts in MMoE are concentrating on new users. That actually causes redundancy. On the contrary, 
POSO solves the problem by assigning personalized modules.
With our POSO, activations, heads and experts are specialized in various users, they become field-aware (see Sec.\ref{sec:vis}). Combined POSO provide significant improvement on both user groups and all the tasks, such improvement further empowers online gains by a large margin.

\begin{table*}[]
\centering
\begin{tabular}{c|cccc|cccc}
\toprule
 &
  \multicolumn{4}{c|}{New Videos} &
  \multicolumn{4}{c}{Regular Videos} \\ \cline{2-9} 
\multirow{-2}{*}{Methods} &
  Long View &
  Finish View &
  Like &
  Follow &
  Long View &
  Finish View &
  Like &
  Follow \\ \hline
  POSO (MLP) &
  \cellcolor[HTML]{FFF0F5}-0.042 &
  \cellcolor[HTML]{FFF0F5}+0.003 &
  +0.292 &
  +0.388 &
  \cellcolor[HTML]{FFF0F5} +0.022 &
  \cellcolor[HTML]{FFF0F5} -0.001 &
   +0.019 &
   +0.003 \\
POSO (MHA) &
  \cellcolor[HTML]{FFF0F5}+0.256 &
  \cellcolor[HTML]{FFF0F5}\textbf{+0.211} &
  +0.520 &
  +0.428 &
  \cellcolor[HTML]{FFF0F5}+0.460 &
  \cellcolor[HTML]{FFF0F5}+0.288 &
  +0.046 &
  +0.442 \\
POSO (MMoE) &
  \cellcolor[HTML]{FFF0F5}+0.163 &
  \cellcolor[HTML]{FFF0F5}+0.164 &
  +0.518 &
  \textbf{+1.456} &
  \cellcolor[HTML]{FFF0F5}-0.059 &
  \cellcolor[HTML]{FFF0F5}-0.009 &
  +0.218 &
  +0.053 \\
POSO (All Combined) &
  \cellcolor[HTML]{FFF0F5}\textbf{+0.269} &
  \cellcolor[HTML]{FFF0F5}\textbf{+0.211} &
  \textbf{+0.727} &
  +0.430 &
  \cellcolor[HTML]{FFF0F5}\textbf{+0.466} &
  \cellcolor[HTML]{FFF0F5}\textbf{+0.298} &
  \textbf{+0.399} &
  \textbf{+0.503} \\ \bottomrule
\end{tabular}
\caption{Offline experiment results on video cold start, compared with baseline.}
\label{tab:offline_newvideo}
\end{table*}

Moreover, we verify whether the proposed method can be adopted to other tasks, such as item cold start (The definition of new video is given in Sec.\ref{sec:coldstart}). To this end we replace the PC by video ages (time from the video uploaded). In Table.\ref{tab:offline_newvideo} we show the results compared with the baseline. 
There are two interesting results: 1) New video POSO performs better results on regular evaluation. 2) It produces larger improvement on regular video samples instead of new video samples. We analyze the results, and figure that the reason is two-fold: on one hand, the system has been struggling for ensuring that new videos can obtain impressions, which essentially sacrifices performance of other videos. On the other hand, existing modules tend new videos excessively. The proposed POSO decouples new/regular videos with their specific modules, thus improves both groups consistently. 
POSO (MLP) improves Like/Follow Rate, but provides competitive results on watching tasks.
It implies value matrices/experts are more capable than activations to balance various user groups.


\begin{table*}
\begin{tabular}{c|c|ccccc}
\toprule
\multicolumn{2}{c|}{Metrics} & Retention Rate & Watch Time &  Like Rate & Follow Rate & Maturing Rate\\ \hline  
\multicolumn{2}{c|}{New User}   &  \color[HTML]{FE0000} \textbf{+1.52\%}   &   \color[HTML]{FE0000} \textbf{+7.75}\%  &    +2.09\%  &   \color[HTML]{FE0000}{\textbf{+11.56\%}}  & - \\ \hline
 \multirow{2}{*}{Inactive User} & active 2 days in past 7 days & - &   \color[HTML]{FE0000} \textbf{+1.50}\% & - & - & -\\
  & active 1 days in past 7 days & - & \color[HTML]{FE0000} \textbf{+1.98\%} & - & - & -\\ \hline
 \multicolumn{2}{c|}{Returning User} & - & \color[HTML]{FE0000} \textbf{+3.01\%} & - & - & -\\ \hline
 \multicolumn{2}{c|}{Regular User} & +0.15\% & \color[HTML]{FE0000} \textbf{+1.99\%} & +0.05\% & +1.42\% \\ \hline 
 \multicolumn{2}{c|}{New Video} & - & +3.81\% & - & - & \color[HTML]{FE0000} \textbf{+0.58\%}\\
 \bottomrule
\end{tabular}
\caption{Online A/B results of various user groups: New User, Inactive user, Returning User, Regular User and New Video. We omit some results because they are undefined or too sparse to give statistical advice. Red results mean they are statistically significant (see text for details). All user groups except new video share the same PC (AVC). }
\label{tab:online}
\end{table*}

\subsection{Online Experiments}
In this section, we conduct online A/B experiments and show the results of large-scale industrial recommender system. We focus on the following metrics (from high importance to low): Watch Time, Retention Rate and Like/Follow Rate.
Watch Time reflects how users are attracted by the recommended videos, and
Retention measures whether a user will keep using the application on the following days. 

We show the online results on various user groups in Table.\ref{tab:online}. Inactive users refer to the users who only keep active few days during a week, and returning users' last launch happens at least $7$ days ago.
Red results mean they are statistically significant, identified by the platform. We view baseline and the experiment as individual distributions. The performance accumulated over a natural day forms a single sample. Then we apply the Student's t-test on the samples. If the experiment samples cannot be generalized by the baseline distribution with more than 95\% probability, the experiment is marked as significant.

First, we discuss the performance on new users. Their Watch Time is significantly improved by 7.75\%. 
Such improvement not only verifies the effectiveness of the proposed method,
it also brings further positive feedbacks to the whole system: new users play more videos and meanwhile their features and training samples are enriched. 
The Follow Rate is improved by 11.56\%, which means the model makes more accurate predictions on user-author relationships. All of the improvement leads to positive results on Retention Rate. We can confirm that new users are more interested in recommended videos than before, and we are more likely to increase DAU (Daily Active Users). For regular users, our method also obtains consistent improvement on Watch Time (+1.99\%), while keeping competitive Retention Rate and interaction metrics. 

One interesting observation is that 
inactive/returning users get significantly improved even that they have been deeply silent. With the silence increases, our model produces larger improvement: from +1.50\%, +1.98\% to +3.01\%.
Combining these results with the ones on regular users and new users, we can conclude that when users tend to be inactive, their distributions deviate and personalization becomes critical. 

We also show the results for video cold start. The metrics are ``Maturing Rate" and per-play Watch Time. The former one describes the speed that new videos become regular ones as defined in Sec.\ref{sec:coldstart}, and the latter averages Watch Time on videos.
POSO significantly improves the primary metric: Maturing Rate. 

In summary, POSO is verified to be effective, and generalized for the large-scale industrial recommender system. It reinforces personalization and improves cold start problem significantly.

\begin{table}[]
\begin{tabular}{c|cc}
\hline
Methods & Favorite  & Satisfied   \\ \hline
Baseline    &    76.08    &     74.57   \\
MeLU~\cite{melu}    &   76.16     &   74.51     \\
Meta-E~\cite{metae}  &   76.13     &   74.53     \\
MWUF~\cite{mwuf}    &   76.20     &   74.51     \\
\hline
POSO    & \textbf{76.89} & \textbf{75.29} \\ \hline
\end{tabular}
\caption{Results on the MovieLens 20M dataset.}
\label{tab:publicdata}
\end{table}

\subsection{Public Dataset}
We verify our method on the public dataset: MovieLens 20M~\cite{movielens}, which collects user rating scores on movies. It contain more than $130k$ users and more than $20$ millions of samples. Since there is no off-the-shelf setup on new user tasks, we split the dataset based on user ID. $100k$ users are divided as training set and the rest is test set. We setup two tasks: 1) Whether the user rates the movie by score=5 (Favorite), and 2) Whether the rating score is not less than 4 (Satisfied). We use two kinds of list features: user's past rating movie ID list and user's past rating tag list, whose length is limited by $30$. The PC is AVC. In both task we use GAUC as the metric.

The results are shown in Table.\ref{tab:publicdata}, existing approaches mostly trade off the performance. Favorite can be improved, however, Satisfied drops. It seems meta-learning methods suit the task with denser positive samples. 
The proposed POSO reaches the best results on both tasks. Interestingly, on a more difficult task (Favorite), it brings larger improvement (0.81pp vs 0.72pp).

\subsection{Evolution of Personalization Code}\label{sec:gates}
In POSO, personalization derives from the usage of Personalization Code (i.e. the input feature of gating network). There could be various choices for the specific designing and formulation of PC. In this section, we study the evolution of PC in the user cold start scenario.

The comparison is shown in Table.\ref{tab:personalizationcode} (measured on new user). The first knowledge is that highlighting personalized feature achieves the effectiveness of POSO. When using all features as input, which degrades into MoE~\cite{moe13,moe17}, we obtain inferior results. That is to say, involving overall features in gating network even worsens submergence.
As for personalized features,
the most trivial choice is an indicator $\vmathbb{1}_{is-new-user}$ whose value equals to $1$ when the user is a new visitor and $0$ otherwise. 
Such PC has provided large-margin improvement.
Since ID embedding implicitly encodes personalization cues, we exploit it as PC. User ID provides moderate improvement. However, its personalization is heterogeneous for cold start task so the results are inferior to the previous PC.
Similarly, Adding video ID embedding further draws back the performance.
The best results are produced by the bucketized Accumulated View Count, which 
counts each impression from the user's first launch, and finely
describes user activity and the period of lifecycle. Its improvement even exceeds the difference between with/without PC.
\begin{table}[]
\begin{tabular}{c|cc}
\toprule
Personalization Code         & Long-View & Finish-View \\ \hline
all features (MoE)           & -0.274          & -0.501          \\
is-new-user                  & +0.240          & +0.145          \\
user ID embedding            & +0.154          & +0.007          \\
user ID + video ID embedding & +0.235          & +0.117          \\
Accumulated View Count       & \textbf{+0.442} & \textbf{+0.248} \\
\bottomrule
\end{tabular}
\caption{Comparison on various Personalization Codes. All results show difference compared with baseline.}
\label{tab:personalizationcode}
\end{table}

\begin{table}[]
\begin{tabular}{c|cc}
\hline
Settings     & Long-View  & Finish-View \\ \hline
$Q^{(i)}$, $K^{(i)}$, $V^{(i)}$, $N=4$  & +0.130 & +0.131   \\
$Q^{(i)}$=$Q$, $K^{(i)}$, $V^{(i)}$, $N=4$ & +0.157 & \textbf{+0.296}   \\
$Q^{(i)}$=$Q$, $K^{(i)}=G^k \odot K$,$V^{(i)}$, $N=4$ & \textbf{+0.279} & +0.190   \\
$Q^{(i)}$=$Q$, $K^{(i)}=G^k \odot K$,$V^{(i)}=G^v \odot V$ & +0.044 & +0.119 \\ \hline


\end{tabular}
\caption{POSO (MHA) under various settings from full personalization to most light-weighted implementation. For simplification we only show new user metrics.}
\label{tab:mhas}
\end{table}

\subsection{Up to What Extent of Personalization?}\label{sec:degree}
In the derivation of POSO we actually have many choices to simplify or maintain the original formulation. Here we take MHA for example, detail performance of each version and explain why we choose the formulation as stated in Sec.\ref{sec:pmom}. 

As shown in Table.\ref{tab:mhas}, the original version of POSO (Eq.\ref{eq:pmom-final}) can already bring better results. However, it cost huge overheads since no simplification is considered. Interestingly, fixing $Q$ provides larger improvement, which also verifies that $Q$ has been highly personalized. Redundantly personalizing it in contrast draws back the performance. Masking $K$ by an element-wise multiplication trades off Long-View Rate and Finish-View Rate. Considering this setting significantly saves computational overheads and meanwhile provides promising results, we use it as the standard POSO (MHA). Further simplifying $V^{(i)}$ reduces the improvement, which degrades personalized $V^{(i)}$ into personalized activations.
As discussed above, $V^{(i)}$ and experts are more capable than activations.

\subsection{Specialization of Modules}\label{sec:vis}
\begin{figure}[t]
  \centering
  \includegraphics[width=0.9\linewidth]{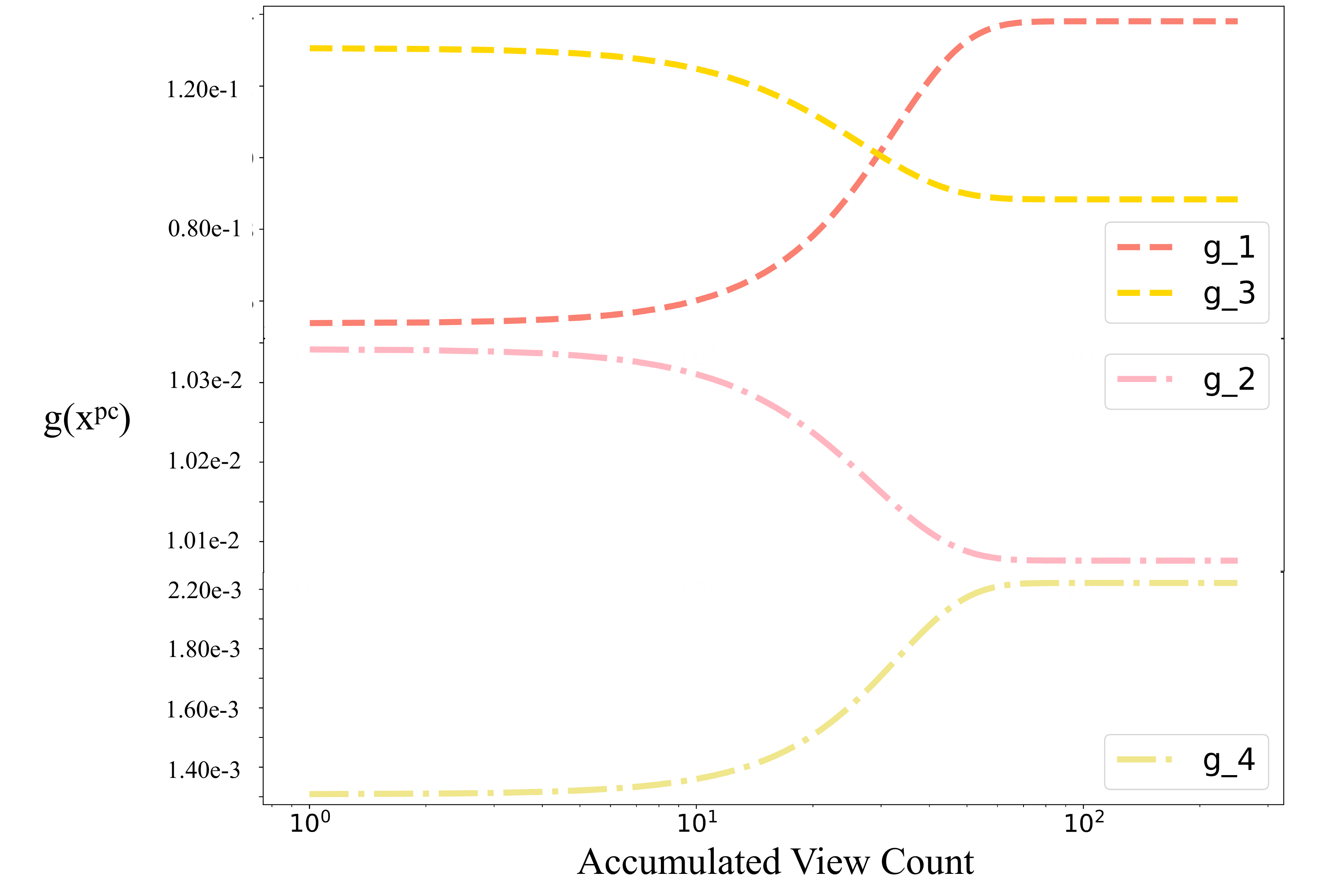}
  \caption{The gating network outputs with bucketized Accumulated View Count increases. \#2 and \#3 are specialized in new users while the others manage regular users. \#1 and \#3 dominate in the combination while \#2 and \#4 work on fine-tuning.}
  \label{fig:vis}
\end{figure}

As a whole, POSO reinforces personalization. Specifically, sub-modules are specialized.
In Fig.\ref{fig:vis} we visualize the gating network outputs for $V^{(i)}$ in our POSO (MHA), which is determined by input (bucketized Accumulated View Count). For new users (lower AVC), gate \#3 is decisive. With AVC increases, gate \#3 gradually becomes underprivileged and gate \#1 dominates. It implies that \#3 has been specialized in managing new users
and \#1 focuses on regular users. \#2 and \#4 performs similarly, however, they work differently and finely tune the final results.

\section{Conclusion}
Personalization is critical for rank model in recommender system. In this paper, we figure that in existing model architecture, imbalanced-personalized features can be easily overwhelmed. To balance various user groups, we propose the Personalized Cold Start Modules method, which flexibly adopts existing methods and derives their personalized versions with negligible computational overheads. The method is verified to effectively improves the performance for new users, new items and returning/inactive users by a large margin. We also discuss the choice of Personalization Code and how to efficiently personalize a specific module. We believe the practical experience can be well generalized to many other scenarios.

\bibliographystyle{ACM-Reference-Format}
\bibliography{pmom}










\end{document}